\documentstyle[aps,pra,epsfig,amsmath,amssymb,twocolumn]{revtex}
\begin{document}
\title{Dynamics of a classical gas including dissipative and mean field
effects}

\author{P. Pedri$^{1,2}$, D. Gu\'ery-Odelin $^{3}$ and S. Stringari $^{1}$}
\address{(1) Dipartimento di Fisica, Universit\`a di Trento
and BEC-INFM, I-38050 Povo, Italy}
\address{(2) Institut f\"ur Theoretische Physik, Universit\"at
Hannover, D-30167 Hannover,Germany}
\address{(3) Laboratoire Kastler Brossel, Ecole Normale Sup\'erieure, 24, Rue
Lhomond, F-75231 Paris Cedex 05, France}

\maketitle

\begin{abstract}
By means of a scaling ansatz, we investigate an approximated
solution of the Boltzmann-Vlasov equation for a classical gas. 
Within this framework, we derive the
frequencies and the damping of the collective oscillations of a
harmonically trapped gas and we investigate its expansion after 
release of the trap. The method is well suited to studying
the collisional effects taking place in the system and in particular
to discussing the crossover between the hydrodynamic and the
collisionless regimes. An explicit link between the relaxation
times relevant for the damping of the collective oscillations and
for the expansion is established.
\end{abstract}


The favored signature of Bose-Einstein condensation in weakly
interacting gases is the time-of-flight expansion \cite{exp95}. In
this technique, the asymmetric trapping potential is switched off
and the evolution of the spatial density is monitored. After long
time expansion, the observed inversion of the aspect ratio
reflects the anisotropy of the initial confinement. In an ideal
Bose-Einstein condensate (BEC), this effect is a direct
consequence of the Heisenberg uncertainty constraint on the
condensate wave function. For an interacting BEC, the inversion is
also produced by the anisotropy of the pressure gradients caused
by the hydrodynamic forces. The changes in the shape of the
expanding gas can be characterized by scaling factors, which
provide an easy quantitative tool for the analysis of on-going BEC
experiments. The set of equations for those factors have been
derived in many papers \cite{scaling,shlyap}. A similar effect has 
also been predicted for a Fermi gas in its superfluid phase
\cite{fermi}. Strong anisotropy has been recently measured in the
expansion of a highly degenerate Fermi gas \cite{thomas} close to a
Feshbach resonance. Resonance scattering can also give rise to
anisotropic expansion in the normal phase as proven in the
experiments of \cite{christophe,djin} carried out in a less
degenerate regime. Some recent experiments on bosonic atoms above
the critical temperature have also reached the collisional regime
investigating both the oscillations of the low lying quadrupole
mode and the expansion in asymmetric traps \cite{HDm,expexp}.

So far analytic calculations for the expansion of a classical gas
have been limited to either the ballistic or to  the hydrodynamic
regime \cite{shlyap}. As a consequence, it is important to generalize such
calculations to all intermediate collisional
regimes. This is precisely the main purpose of this paper. We
begin by an outline of the theoretical description of the thermal
gas based on the Boltzmann-Vlasov equation. Our approach relies on
an approximated solution to this equation by means of a scaling
ansatz. This solution is used throughout the paper to investigate
two kinds of related problems: the lowest collective oscillation
modes and the time-of-flight expansion when the confinement is
released.

\noindent

The Boltzmann-Vlasov (BV) kinetic equation for the phase space
distribution $f(t,{\bf r},{\bf v})$ takes the form:

\begin{eqnarray}
\frac{\partial f}{\partial t}+ {\bf v}\cdot\frac{\partial
f}{\partial {\bf r}}-\frac{1}{m}\frac{\partial (U_{\rm h}+U_{\rm
mf})}{\partial {\bf r}} \cdot\frac{\partial f}{\partial {\bf
v}}=I_{\rm coll}[f], \label{BV}
\end{eqnarray}
where $U_{\rm h}$ is the trapping potential chosen of harmonic
form: $U_{\rm h}({\bf r})=\frac{m}{2}\sum_i \omega_i^2r_i^2$.
Interparticle interactions enter Eq. (\ref{BV}) in two different
ways \cite{griffin}. On the one hand, they modify the effective
potential through the mean field term $U_{\rm mf}$  which affects
the streaming part of the Boltzmann kinetic equation. The mean
field potential  $U_{\rm mf}$ is equal to $2gn$ for bosons
 and $gn$ for two fermion species \cite{fermnot}, where
the coupling constant $g=4\pi \hbar^2a/m$ is fixed by the $s$-wave
scattering length $a$.
 The mean field
term is linear in $a$ and is non dissipative. On the other hand,
two body interaction determines the collision integral $I_{\rm
coll}[f]$ which is quadratic in the scattering length, and
describes dissipative processes. Eq. (\ref{BV}) is valid in the
semi-classical limit, namely when the thermal energy is large
compared with the separation between the energy eigenvalues of the
potential \cite{Huang,BaymLivre}.

In this paper, we will treat the collision integral within the
relaxation time approximation. This model should suffice to
capture the essential physics of the problem. We consequently
write:
\begin{eqnarray}
I_{\rm coll}[f]\approx-\frac{f-f_{\rm le}}{\tau},
\end{eqnarray}
where $\tau$ is the relaxation time related to the average time
between collisions, and $f_{\rm le}$ is the local equilibrium
density in phase space. As a consequence, $f_{\rm le}$
 has a spherical symmetry in velocity space,
{\it i.e.} it depends on the velocity through $[{\bf v}-{\bf
u}({\bf r})]^2$ where ${\bf u}(\bf r)$ is the local velocity
field.

The dynamics of the gas will be described by the
following scaling ansatz for the non equilibrium distribution
function:
\begin{eqnarray}
f(t,r_i,v_i)= \frac{1}{\prod_j(b_j\theta_j^{1/2})}
f_0\left(\frac{r_i}{b_i},
\frac{1}{\theta_i^{1/2}}\left(v_i-\frac{\dot{b}_i}{b_i}r_i\right)\right),
\end{eqnarray}
where $f_0$ is the equilibrium distribution function which
satisfies the equation ($I_{\rm coll}[f_0]=0$):
\begin{eqnarray}
\label{beq} m{\bf v}\cdot\frac{\partial f_0}{\partial {\bf r}}=
\frac{\partial U_{\rm h}}{\partial {\bf r}}\cdot\frac{\partial f_0
}{\partial {\bf v}} +\frac{\partial U_{\rm mf}}{\partial {\bf r}}
\cdot\frac{\partial f_0}{\partial {\bf v}}.
\end{eqnarray}
The dependence on time of $f$ is contained in the dimensionless
scaling parameters $b_i$ and $\theta_i$. The parameter $b_i$ gives
the dilatation along the $i^{th}$  direction, while $\theta_i$
gives the effective temperature  in the same direction. Such an
ansatz generalizes the one used in \cite{guer}. We recall that in
this method
 the shape of the cloud does not explicitly enter the
 equations. This is the reason why the solutions are equally
 valid for a dilute Bose gas above the critical temperature, a dilute Fermi gas
in its normal
 phase,
  and a classical gas.

Following Ref. \cite{guer}, one can derive the set of equations
for the scaling parameters $b_i$ and $\theta_i$ (see Appendix A):
\begin{eqnarray}
&&
\ddot{b}_i+\omega_i^2b_i-\omega_i^2\frac{\theta_i}{b_i}+\omega_i^2\xi
\left( \frac{\theta_i}{b_i} -\frac{1}{b_i\prod_j b_j}\right)=0
 \label{s1}\\
&& \dot{\theta}_i+2\frac{\dot{b}_i}{b_i}\theta_i=-\frac{1}{\tau}
\bigg[\theta_i-\bar{\theta}\bigg], \label{s2}
\end{eqnarray}
where the dimensionless parameter $\xi=\langle U_{\rm mf} \rangle_0/(\langle
U_{\rm mf} \rangle_0+2m\langle v^2 \rangle_0/3)$ accounts for the
mean field interaction \cite{xin} and $\bar{\theta}=\sum_i
\theta_i/3$ is the average temperature, for a classical gas $\langle
v^2\rangle_0=3k_BT/m$. The parameter $\xi$ is
expected to be small for dilute gases ($na^3\ll 1$) since the
ratio $U_{\rm mf}/k_BT$ scales as
$(na^3)^{1/3}(n\lambda_{db}^3)^{2/3}$, where $\lambda_{db}$ is the
de Broglie wavelength and $n$ the mean density \cite{guer}. Eq.
(\ref{s2}) shows that the dissipation occurs when the temperature is not
isotropic and the relaxation time $\tau$ has a finite value.

Eqs (\ref{s1}) and (\ref{s2}) are the main result of this paper.
The collisionless regime is obtained by taking $\tau_0=\infty$. In
this limit, we have the simple relation $\theta_i=b_i^{-2}$
between the scaling parameters, and we recover the equations
derived in \cite{guer}. In the opposite limit (hydrodynamic
regime), local equilibrium is always ensured because of the high
collision rate. As a consequence, we have
$\theta_i=\bar{\theta}=\prod_j b_j^{-2/3}$ and the Eqs (\ref{s1})
and (\ref{s2}) can be recast in the form:
\begin{eqnarray}
&&\ddot{b}_i+\omega_i^2b_i-\frac{\omega_i^2}{b_i\prod_j b_j^{2/3}}+
\nonumber\\
&& \omega_i^2\xi \left( \frac{1}{b_i\prod_j b_j^{2/3}}
-\frac{1}{b_i\prod_j b_j} \right)=0.
\end{eqnarray}
For $\xi=0$ (no mean field), we recover the equations first
derived in \cite{shlyap}. Note that in both the collisionless and
the hydrodynamic regimes, the collisional term does not contribute
since there is no dissipation in these limits. We next focus our
attention on the intermediate regimes where the collision term
enters explicitly the equations of motion.

Let us first study the breathing mode in the case of  spherical
harmonic trapping with angular frequency $\omega_0$. In this case,
we find a solution with $b_i=b$ and $\theta_i=b^{-2}$. For such a
solution the collision term identically  vanishes in all
intermediate collisional regimes. Our approach can be readily
generalized to lower dimensions leading to the frequency $\omega_0
(4+\xi(d-2))^{1/2}$ for the monopole mode \cite{guer}, where $d$
is the dimension of space. In two dimensions the mean field does
not affect the frequency of the monopole. This comes out from the
fact that in this case the ansatz is an exact solution of the BV
equations, as already stressed  in Ref. \cite{rosch}.

We now consider a sample of atoms confined in a three-dimensional 
cylindrically symmetric harmonic potential. We denote by
$\lambda=\omega_z/\omega_\perp$ the ratio between the axial
 and radial angular frequencies. Expanding Eqs. (\ref{s1}) and
(\ref{s2}) around equilibrium ($b_i=\theta_i=1$) we get a linear
and closed set of equations which can be solved by looking for
solutions of the type $e^{i\omega t}$. The associated determinant
yields the dispersion law:
\begin{equation}
\bigg(A[\omega]+\frac{1}{\tau_0}B[\omega]\bigg)
\bigg(C[\omega]+\frac{1}{\tau_0}D[\omega]\bigg)=0, \label{disp}
\end{equation}
where $A[\omega]=\omega^2(\omega^2-\omega_{\rm cl
+}^2)(\omega^2-\omega_{\rm cl-}^2)$,
$B[\omega]=\omega(\omega^2-\omega_{\rm hd
+}^2)(\omega^2-\omega_{\rm hd -}^2)$,
$C[\omega]=\omega(\omega^2-\omega_{\rm cl }^2)$ and
$D[\omega]=(\omega^2-\omega_{\rm hd}^2)$ and $\tau_0$ is the value
of the relaxation time $\tau$ calculated at equilibrium and 
\begin{eqnarray}
\omega^2_{\rm cl \pm} & = &
\frac{\omega^2_\perp}{2}\bigg(4(1+\lambda^2)-\lambda^2\xi
\nonumber
\\ && \pm \sqrt{16+\lambda^4(4-\xi)^2+8\lambda^2(\xi^2-4+\xi)}\bigg)
\nonumber
\\
\omega^2_{\rm cl} & = & \omega^2_\perp(4-2\xi) \nonumber
\\
\omega^2_{\rm hd \pm} & = & \frac{\omega^2_\perp}{3}
\bigg(5+4\lambda^2+\xi(1+\lambda^2/2)\nonumber\\&& \pm
\frac{1}{2}\sqrt{(10+8\lambda^2+2\xi+\lambda^2\xi)^2-72\lambda^2(4+\xi)}\bigg)
\nonumber
\\
\omega^2_{\rm hd} & = & 2\omega^2_\perp. \nonumber
\end{eqnarray}
Here $\xi$ is the parameter accounting for the mean field effects and (cl)
and (hd) refer to the collisionless and hydrodynamics regimes respectively.
The solution of Eq. (\ref{disp}) interpolates the frequencies of
the low lying modes for all collisional regimes ranging from the
collisionless to the hydrodynamic. As the confinement is
cylindrically symmetric around the $z$ axis, we can label the
modes by their angular azimuthal number $M$. The first factor of
the l.h.s. of Eq. (\ref{disp}) gives the frequencies of the two
$M=0$ modes, while the second factor gives that of the quadrupole
($M=\pm 2$) mode. The roots of $A$ and $C$ have already been obtained in
\cite{guer}, and correspond to the frequencies of the low lying
modes of a collisionless gas in presence of mean field. Eq.
(\ref{disp}) for $\xi=0$ has been derived in Ref. \cite{dgo99} and the
corresponding frequencies have been 
investigated experimentally \cite{HDm}. For $\xi=1$, corresponding
to $\langle gn_0\rangle_0 \gg \langle v^2\rangle_0 $, we find
$\omega^2_{\rm cl \pm}=\omega^2_{\rm hd \pm}$, $\omega^2_{\rm
cl}=\omega^2_{\rm hd}$, and the frequencies coincide with the ones predicted 
for a Bose-Einstein condensate in the Thomas-Fermi regime
\cite{colex}. 

So far, we have not given the explicit link between
the relaxation time entering Eq.~(\ref{disp}) and the collision rate. Following
Ref. \cite{dgo99}, we can establish this link for a classical gas
by means of a Gaussian ansatz for the equilibrium distribution
function $f({\bf r},{\bf v},t)$. One obtains $\tau_0=5/(4\gamma)$
where $\gamma=2(2\pi)^{-1/2}n_{\rm max}\sigma v_{\rm th}$ is the
classical collision rate where $v_{\rm
th}=(k_BT/m)^{1/2}$ is the thermal velocity, 
$n_{\rm max}$ is the peak density and
 $\sigma$ is the cross section which is assumed to be velocity independent. 
 For bosons the link between the scattering length and the cross 
 section is $\sigma=8\pi a^2$ whereas for two fermions species 
 one has $\sigma=4\pi a^2$.

We now establish the set of equations that describe the
time-of-flight expansion. In the collisionless regime where the
mean free path is very large with respect to the size of the
trapped cloud and in the absence of mean field contribution, we
readily obtain the exact equations $\ddot{b}_i=\omega_i^2/b_i^3$
which admit the solutions $b_i(t)=(1+\omega_i^2t^2)^{1/2}$,
leading to isotropic density and velocity distributions after long
time expansion.

When the effect of collisions is important the physics of the
expansion changes dramatically. As an example, the radial
directions of a cigar-shaped cloud expand faster than the
longitudinal one finally resulting in an anisotropic velocity
distribution. So far, an analytic approach has been proposed only
in the full hydrodynamic regime \cite{shlyap}. However, this
approach assumes that the hydrodynamic equations are always valid
during the expansion. In general, this cannot be the case since
the density decreases during the expansion reducing the effect of
collisions. Alternatively, the expansion of an interacting Bose above $T_c$
gas has been investigated by means of a Monte Carlo simulation
 \cite{ARIMONDO}.

In our approach, we provide an interpolation between the two
opposite collisionless and  hydrodynamic regimes using the scaling
formalism. The decrease of the collision rate during the expansion
yields a non constant relaxation time $\tau(b_i,\theta_i)$ that depends 
explicitly on the scaling parameters and reflects the changes
of the density and the temperature during the expansion. As a
result, the expansion is described by the following set of 6
non-linear equations
\begin{eqnarray}
\ddot{b}_i-\omega_i^2\frac{\theta_i}{b_i}+ \omega_i^2\xi \left(
\frac{\theta_i}{b_i}
-\frac{1}{b_i\prod_j b_j} \right)=0\nonumber\\
\dot{\theta}_i+2\frac{\dot{b}_i}{b_i}\theta_i=-\frac{1}{\tau(b_i,\theta_i)}
\left( \theta_i-\frac{1}{3}\sum_j\theta_j \right). \label{expan}
\end{eqnarray}
The dependence of the relaxation time $\tau$ on the scaling parameters is
obtained by
noticing that the collision rate $\gamma$ scales as $n T^{1/2}$. Using the
scaling transformation $n\rightarrow n_0(\prod_j b_j)^{-1}$ and $T\rightarrow
T_0\bar{\theta}$, where $n_0$ and $T_0$ are the initial density and temperature
respectively, we deduce
\begin{equation}
\tau(b_i,\theta_i)=\tau_0\bigg(\prod_jb_j\bigg)\bigg(\frac{1}{3}
\sum_k\theta_k\bigg)^{-1/2}
\label{tauexp}
\end{equation}
where $\tau_0$ is the average time of collision at equilibrium
\cite{unilim}. Since both the results  (\ref{disp}) for the
dispersion of the linear oscillations and Eqs. (\ref{expan}) and
(\ref{tauexp}) for the expansion have been derived starting from
the same scaling equations  (\ref{s1}) and (\ref{s2}) the
relaxation time $\tau_0$ entering the two processes is the same.
As a consequence, the combined investigation of the expansion and
of the quadrupole oscillations can provide a useful check of the
consistency of the approach and, possibly, useful constraints on
the value of the cross section.

The time evolution of the aspect ratio $R_\perp(t)/R_z(t)=\lambda
b_\perp(t)/b_z(t)$ in the absence of mean field is depicted on
Fig. 1 for different values of the product $\omega_\perp \tau_0$
and for an initial aspect ratio $\lambda=0.1$. In the
collisionless regime ($\tau_0=\infty$), the aspect ratio tends
asymptotically to unity reflecting the isotropy of the initial
velocity distribution. For other collisional regimes, the
asymptotic aspect ratio is larger that one as a consequence of
collisions during the expansion. We find a continuous transition
from the collisionless to the hydrodynamic prediction as we
decrease $\tau_0$ from infinity to zero. We then conclude that in
general the expansion cannot be described with either the
hydrodynamic or the collisionless prediction \cite{walra}, but
requires a full solution of our equations (\ref{expan}). Similar
conditions have been already encountered experimentally
 for classical or almost classical gases \cite{expexp}. We also notice that it
 is very important to take into account the time dependence of the relaxation
 time, accounted for by the scaling law (\ref{tauexp}). 
 For example by simply using $\tau=\tau_0$ during the whole expansion, the
curve
$\omega_\perp\tau=0.1$ of Fig. 1 (solid line) would be shifted upward and the
resulting prediction would result much closer to the hydrodynamic curve 
(dotted line).

\begin{figure}
\begin{center}
\epsfig{file=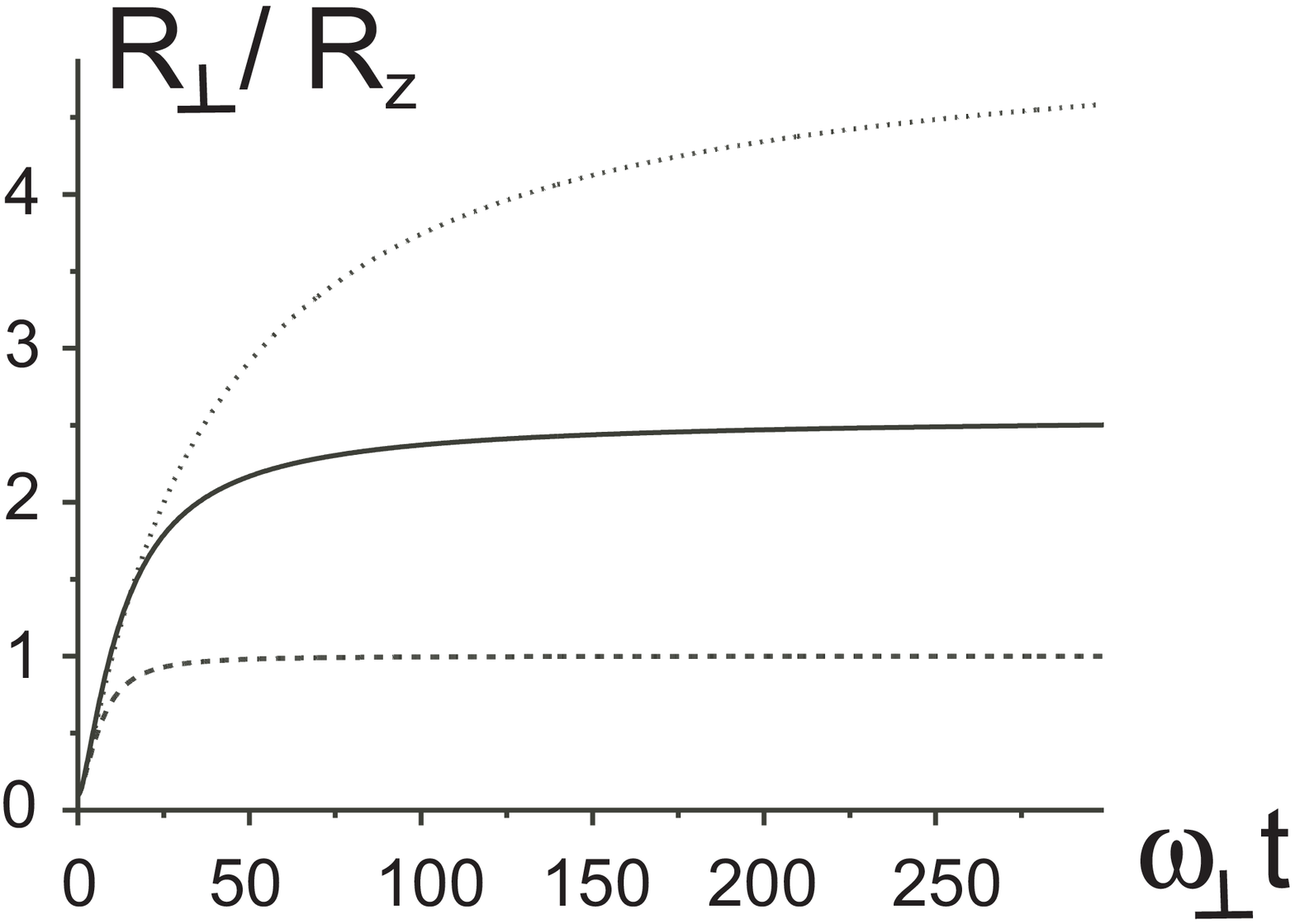, width=8cm}
\begin{caption}
{\sl Aspect ratio as a function of the normalized time
$\omega_\perp t$ for different collisional regimes (initial aspect
ratio $\lambda=0.1$): collisionless (dashed line,
$\tau_0\longrightarrow \infty$), intermediate collisional regime
(solid line, $\omega_\perp\tau_0=0.1$), hydrodynamic regime
(dotted line, $\tau_0 \longrightarrow 0$).}
\end{caption}
\end{center}
\label{fig1}
\end{figure}

Let us finally comment on the effect of quantum statistics on the
calculation of the relaxation time. For a Bose gas at temperature
above $T_c$ the problem has been investigated in \cite{ka00} where
it has been shown that statistical effects do not play a
significant role. In contrast, the relaxation time in a
harmonically trapped dilute Fermi gases has been shown to be
strongly affected by Pauli blocking at low temperature
\cite{vichi}. The effects of collisions in a strongly interacting
Fermi gas, including the unitarity limit, have been recently
addressed in \cite{thomasg}.

\section*{Acknowledgments:} This work was supported by the
Bureau National de la M\'etrologie, the D\'el\'egation
G\'en\'erale de l'Armement, and the R\'egion Ile de France, the
Deutsche Forschungsgemeinschaft SFB 407 and SPP1116, the RTN Cold
Quantum Gases, IST Program EQUIP, ESF PESC BEC2000+, EPSRC, and
the Ministero dell'Istruzione, dell'Universit\`a e della Ricerca
(MIUR). P.P. also wants to thank the Alexander von Humboldt Foundation 
and the ZIP Programm of the German government.

\begin{appendix}
\section{Equations}

We make the following ansatz for the non equilibrium distribution
function:  $f({\bf r},{\bf v},t)=\Gamma f_0({\bf R}(t),{\bf
V}(t))$ with $R_i=r_i/b_i$,
$V_i=(v_i-\dot{b}_ir_i/b_i)\theta_i^{-1/2}$ and $\Gamma=\Pi_j
b_j^{-1}\theta_j^{-1/2}$. The dependence on time is contained
through the
 parameters $b_i$ and $\theta_i$. Following \cite{guer}, we substitute
  this ansatz into Eq.
 (\ref{beq}) and use the equation for the equilibrium distribution $f_0$. We
find
\begin{eqnarray}
&&\dot{\Gamma}f_0+\Gamma\sum_i \bigg\{ V_i \frac{\partial
f_0}{\partial
R_i}\bigg(\frac{\theta_i^{1/2}}{b_i}-\frac{1}{b_i\theta_i^{1/2}}\frac{1}{\Pi_j
b_j} \bigg)\nonumber
\\ && -
 \frac{\partial f_0}{\partial
V_i}\bigg[ \frac{R_i}{\theta_i^{1/2}}\bigg(
\ddot{b}_i+\omega_i^2b_i-\frac{\omega_i^2}{b_i}\frac{1}{\Pi_jb_j}
\bigg) \nonumber
\\ &&  +V_i\bigg(
\frac{1}{2}\frac{\dot{\theta}_i}{\theta_i}+\frac{\dot{b}_i}{b_i}\bigg)\bigg]
\bigg\}=I_{\rm
coll} \label{bs}
\end{eqnarray}
Performing integration in phase space, we calculate the average
moment of $R_iV_i$, namely $\int R_iV_i[Eq. (\ref{bs})]
d^3Rd^3V/N$. This leads to Eq. (\ref{s1}). Note that this equation
is not affected by the collision integral since the quantity
$R_iV_i$ is conserved by collisions.

 To
derive Eq. (\ref{s2}), we consider instead the average moment of
$V_i^2$. This yields:
\begin{eqnarray}
\frac{\dot{\theta}_i}{\theta_i}+2\frac{\dot{b}_i}{b_i}=\frac{m}{N\Gamma
k_BT_0}\int  V_i^2 I_{\rm coll}d^3Rd^3V, \label{z1}
\end{eqnarray}
where $T_0$ is the equilibrium temperature. Differently from
(\ref{s1}), Eq. (\ref{z1}) depends explicitly on the collision
integral .
 In order to calculate the r.h.s of Eq. (\ref{z1}), we use the
relaxation time approximation: $I=-(f-f_{\rm le})/\tau_0$. The
first term gives: $\int V_i^2fd^3Rd^3V =  N\Gamma k_BT_0/m$. To
obtain a relation among the $\theta$ scaling parameters one uses
the identity $\langle v^2 \rangle=\langle v^2 \rangle_{\rm le}$,
from which we deduce $\bar{\theta}=\sum_i \theta_i/3$. The
contribution of the local equilibrium term to the second term is
obtained by noticing that, at local equilibrium, $\theta_{i}^{\rm
le}=\bar{\theta}$: $\int V_i^2f_{\rm le}d^3Rd^3V= \bar{\Gamma}\int
V_i^2f_0 (\bar{{\bf R}},\bar{{\bf V}})d^3Rd^3V  =
N\Gamma\bar{\theta}k_BT_0/(m\theta_i)$. Hence Eq. (\ref{z1}) can
be recast in the form (\ref{s2}).
\end{appendix}


\begin{references}
\bibitem{unilim} In the unitary limit of $s$-wave collisions, the cross section $\sigma$
scales as $k^{-2}$ where $k$ is the wave vector for the relative
velocity which suggests a different scaling for $\tau_0$ given by
$\tau_0(b_i,\theta_i)=\tau_0(\prod_jb_j)(\frac{1}{3}\sum_k\theta_k)^{1/2}$.
\end{references}
\end{document}